\newcommand{\beq}{\begin{equation}}
\newcommand{\eeq}{\end{equation}}
\newcommand{\eqna}{\begin{eqnarray}}
\newcommand{\eqne}{\end{eqnarray}}

\documentclass[amsmath,amssymb]{revtex4}
\usepackage{graphicx}
\usepackage{pdfsync}
\usepackage{color}
\begin{document}
\setlength{\voffset}{.5cm}
\title{Analysis of  light scattering off   photonic crystal slabs in terms of Feshbach resonances}
\author{I. Evenor  $^{{\rm a}}$}  \author{E. Grinvald $^{{\rm a}}$}    \email{eran.grinvald@weizmann.ac.il}
\author{F. Lenz $^{{\rm b}}$}    \email{flenz@theorie3.physik.uni-erlangen.de}
\author{S. Levit$^{{\rm a}}$
}    \email{shimon.levit@weizmann.ac.il}
\affiliation{$^{{\rm a}}$ Department of Condensed Matter \\ The Weizmann Institute of Science\\ Rehovot Israel   \\ \\
$^{{\rm b}}$ Institute for Theoretical Physics III \\
University of Erlangen-N\"urnberg \\
Staudstrasse 7, 91058 Erlangen, Germany\\ }
\date{September 11, 2012}
\begin{abstract}
Techniques to deal with  Feshbach resonances   are applied to describe
resonant light scattering off one dimensional photonic crystal slabs.  Accurate  expressions for scattering amplitudes, free of any fitting  parameter,    are obtained for isolated as well as overlapping  resonances. They  relate the resonance properties to the properties  of the optical structure and of the incident light. For the most common case of   a piecewise constant  dielectric function, the calculations  can be carried out essentially analytically. After establishing the accuracy of this approach  we  demonstrate  its potential  in the  analysis of  the reflection coefficients for the diverse shapes of overlapping, interacting  resonances.
\end{abstract}

\maketitle
\section{Introduction}
\begin{figure}[ht] \centering
\vspace{-.05cm}
\includegraphics[width=.5\linewidth]{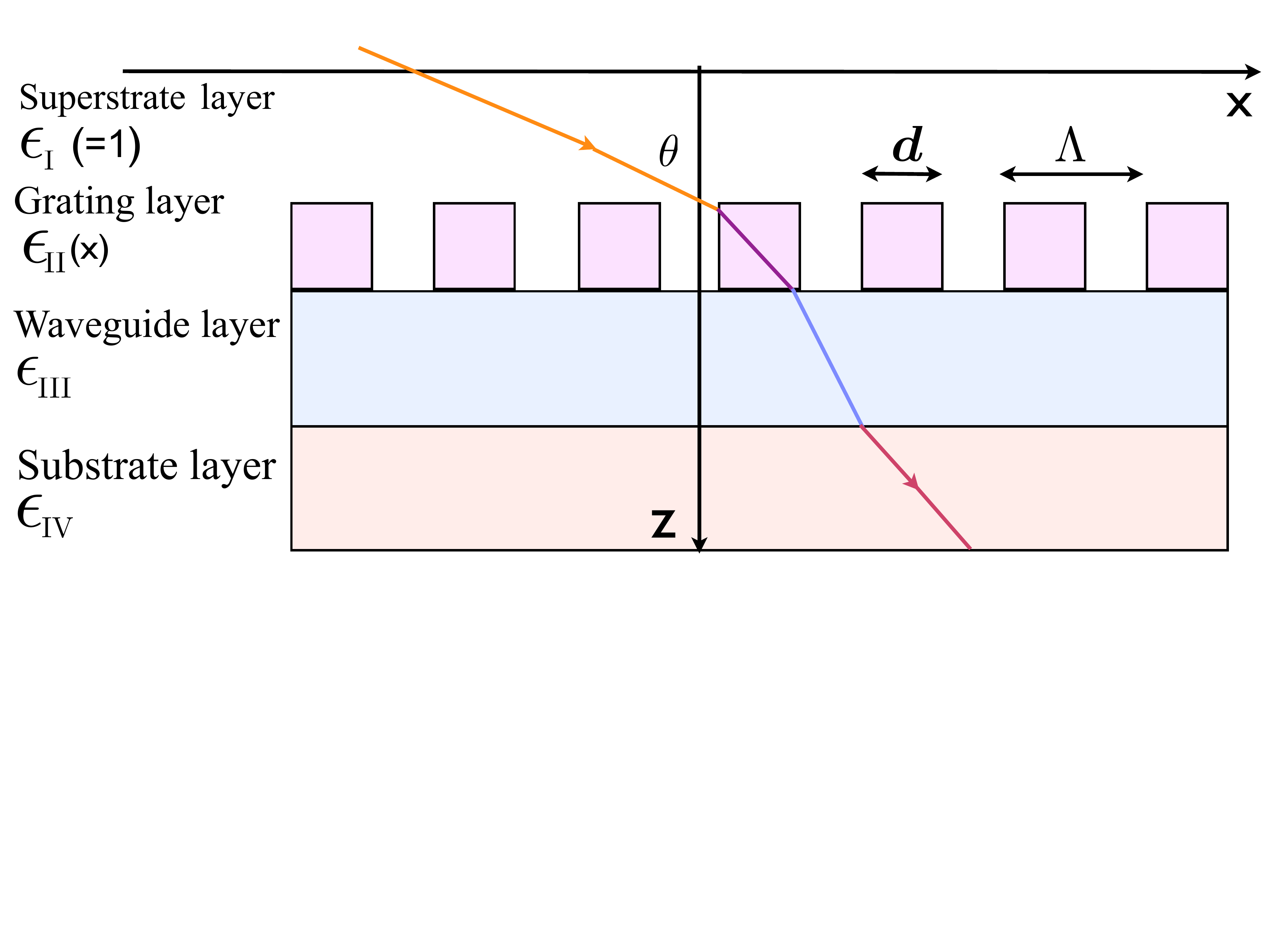}
\vskip -2.15cm
\caption{
Light incident on a photonic crystal slab with piecewise constant layers with dielectric constants $\epsilon_I ...\epsilon_{IV}$. Grating period $\Lambda$ and duty cycle $d/\Lambda$ characterize the grating layer (II).}
\vskip -1cm
\label{dispn}
\end{figure}

The concept of Feshbach resonances \cite{FESH58} has been conceived and applied in the context of  quantum mechanical  scattering  off many-body systems such as  nuclei \cite{FESH92} atoms or molecules\,\cite{TTHK99,CGJT09}.
In these systems a Feshbach resonance occurs if the kinetic energy  of the incident particle is close to an almost stable intermediate ``molecular'' state.
Elegant techniques  to treat   Feshbach resonance scattering  were developed, cf.\,Ref. \cite{FESH58}, leading to  simple and  accurate approximation methods.
In  the present study we propose to extend and apply these techniques to light scattering off photonic crystal slabs. We consider the  case of a one dimensional photonic crystal slab -- the grating waveguide structure (GWS), cf.\,Refs. \cite{WARO907,RSA97}, shown in Fig.\,\ref{dispn}.  The  Feshbach resonances are formed by turning  a ``bound state'' (guided mode) into a resonance ``state'' (barely radiating mode). We will establish the formal connection to Feshbach resonances in quantum  systems and, on this basis, we  will develop a novel, essentially fully analytical approach to GWS resonances.   It is straightforward to generalize this  approach  to resonance scattering off  more complicated, two and three dimensional photonic systems. Its degree of accuracy  in these systems however remains to be established.

As for cold atoms where the  condition for appearance of Feshbach resonances can be tuned by varying the strength of  an external magnetic field,   optical systems offer  a variety of ways to influence appearance and properties of  resonances. By manipulating the parameters of the GWS, qualitative changes of the  shape of  isolated Feshbach resonances  through their Fano interference  with non resonant scattering component  and,  even more interesting, the  generation of two or more overlapping resonances can be attained. Apart  from the conceptual connection with Feshbach resonances in atomic and nuclear systems our analytic approach carries  significant  practical value. It  establishes  easy and intuitively clear relations between the basic photonic crystal slab parameters and the properties of the resonances. It allows  systematic studies of  various phenomena arising as a result of the resonance interaction. It provides analytic tools to design photonic crystal slabs exhibiting light scattering  resonances with desired properties and to understand how to control them.

 Guided mode resonances in GWS were studied in  Refs.\, \cite{MP85}, \cite{GSST85} and are found in a wide variety of applications, cf.\,Refs.\,\cite{KLGY10,WM93,MSJ10,BFCBBDKTS10,FLPFB10,MCWEC08,DTZYJF00,BLTFS09,LTSYM98,KYFMHZ05}.
 In   most theoretical investigations  of these resonances \cite{FMS02,FSJ03,FAJO02,K03}  parameter fitting is required for comparison  with data or exact numerical calculations \cite{MBPK95,SSS82}.   Recently in  Refs. \cite{MFK10,M09,KCLG11} it was recognized  that various photonic resonances can be treated as (asymmetric) Fano-Feshbach resonances and their shapes characterized  by  a fit  to  the Fano formula.   Simple models have been developed, cf.\,\cite{MFK10,ZHTH10} which exhibit optical Feshbach or Fano-Feshbach  resonances.  The majority  of the theoretical approaches  concentrated on the  properties of resonances  with TE polarized light. In  our  formalism  a common treatment of TE and TM resonances arises naturally and provides insights into the origin of their  different properties.

The central results of  our formal development  are the expressions (\ref{tore}),  (\ref{TMr}) for the reflection amplitudes    of   isolated  and  the expressions (\ref{sieps}), (\ref{siepsH}) for overlapping TE and TM resonances respectively. Given the   dielectric function of the   grating waveguide structure  and   wavelength and   angle of the incident light all the quantities appearing in these expressions are  analytically calculable. By comparing  with the ``exact'' numerical results we establish the accuracy of our formulation for  isolated as well as for overlapping resonances.  (For a more detailed presentation of  formalism and results cf.\,\cite{EGLL11}.)

It is important to stress that our method is not limited to scattering amplitudes. It actually  provides analytic expressions for the entire spatial distributions of the electric and magnetic fields  inside and outside the dielectric structure. As an example we use this to analyze  the strength of the resonating electric field  at the end of Section \ref{sec:interact}.
\section{Isolated resonances}
 We shall consider the so called ``classical''  incidence (vanishing y-component of the wavevector)  for which  TE and TM modes decouple and can be treated separately,  cf. Ref.\,\cite{JJWM08}.

In TE polarized waves the electric field  is parallel to the grating grooves. In the  coordinate system of Fig.\,\ref{dispn} its non-vanishing y-component satisfies the  wave equation ($c=1$)
\begin{equation}
 \Big(-\frac{\partial^2}{\partial z^2} -\frac{\partial^2}{\partial x^2} -\epsilon(x,z) \omega^2 \Big)E(x,z)= 0\,,
\label{swva1}
\end{equation}
with $\omega$ denoting the frequency of the incident light.
The non-trivial x-dependence of the  dielectric constant is limited to the  grating layer and is periodic  with  grating period $\Lambda$  and corresponding   reciprocal lattice  vector $K_g=2\pi/\Lambda$. Its Fourier components are given by
\begin{equation*}
\epsilon_n(z)=\frac{1}{\Lambda}\int_{-\Lambda/2}^{\Lambda/2} dx \,\epsilon(x,z) e^{-in K_g x}\,.
\end{equation*}
In terms of the Fourier components $E_n(z)$ of the electric field the wave equation (\ref{swva1}) is
converted into a system of ordinary differential equations
\begin{equation} \label{syseq}
\sum_mD_{nm}E_m(z)=0\,,
\end{equation}
with
\eqna
D_{nm} =  \Big(\frac{d^2}{dz^2} -(k_x+n K_g)^2+\omega^2\epsilon_0(z)\Big)\delta_{nm}+ 
 \omega^2\epsilon_{n-m}(z)(1-\delta_{nm})\,,
\label{DNM}
\eqne
and  the components of the wave vector $k_x,\,k_z$ of the incident light.

Systems of equations of coupled channels with similar structure are underlying the description of Feshbach resonances  in atomic and nuclear physics.
 A Feshbach resonance occurs if, by neglecting  certain  couplings,  a bound state of the entire system, projectile plus target, exists which when  turning on the couplings is converted   into a resonance. We proceed here in the same spirit.

  In analogy with the distinction between ``closed'' and ``prompt'' or ``open'' channels  \cite{FESH92} we  distinguish here coupled evanescent modes $E_{n}(z)$   which for $z\to \pm \infty$  do satisfy the condition
$$\epsilon_0({\pm \infty})\omega^2-(k_x+nK_g)^2 < 0\,,$$  and  extended modes  which do not. To simplify   the  discussion we focus  on the important special case of a  ``subwavelength grating'' where  the kinematics is chosen such that the above  condition is satisfied  only for the $n=0$,  extended mode. In terms of the  (yet undetermined)  evanescent modes, the extended mode is given by
\begin{eqnarray}
E_0(z)= E_0^{(+)}(z)
- \omega^2\int dz^\prime g_0^E(z,z^\prime)\sum_{n\neq 0}\epsilon_{-n}(z^\prime) E_{n}(z^\prime)\,,
\label{extmod}
\end{eqnarray}
as is easily  verified by applying $D_{00}$ to this equation. Here the electric field $E_0^{(+)}$ and the Green's function  $g_0^E$ solve the homogeneous and the inhomogeneous wave equation respectively
\begin{equation}
D_{00}\,E_0^{(+)}=0\,,\quad    D_{00}\,g_0^E(z,z^\prime) = \delta(z-z^\prime)\,,
\label{EGR}
\end{equation}
subject to the boundary conditions imposed by the kinematics illustrated in Fig.\,\ref{dispn}. For $E_0^{(+)}$ they are
\eqna
\lim_{z\rightarrow-\infty} E_0^{(+)}(z)=e^{ik_z z} + r_0^E e^{-ik_z z}\,,  \quad\lim_{z\rightarrow \infty}\hspace{.08cm} E_0^{(+)}(z)=t_0^E  e^{i\tilde{k}_z z}\,,
\label{eq:boundcondTEE}
\eqne
 with  $k_z,\,\tilde{k}_z$ denoting the  z-components of  the wave vector   in the superstrate and  substrate layers  respectively. The electric field  $E_0^{(+)}$   accounts for the   ``background scattering'' of light   off a dielectric medium  with the  $x-$averaged $z-$dependent dielectric constant $\epsilon_0(z)$. The corresponding (``background'') reflection and transmission amplitudes are  denoted by $r_0^E$  and $t_0^E$.  As is well known \cite{MOFS53}, having determined $E_0^+(z)$ together with  it's partner $E_0^-(z)$ (wave incident from the substrate), the  Green's function $g_0^E(z,z^\prime)$ is given by
\begin{eqnarray}
g_0^E(z,z^\prime) = \frac{ E_0^+(z_>)E_0^-(z_<)}{ w^E}\,,\quad
w^E=E_0^+(z)\,dE_0^{-}(z)/dz -E_0^-(z)\,dE_0^{+}(z)/dz\,,
\label{grfcE}
\end{eqnarray}
with $z_{<,>}$ denoting the smaller and larger of the arguments $z,z^\prime$ respectively and $w^E$ the $z$-independent Wronskian.

The evanescent modes in (\ref{extmod}) can in turn be expressed  in terms of $E_0(z)$
\begin{eqnarray}
E_n(z) &=& \sum_m \int dz^\prime\big(\tilde{D}^{-1}\big)_{nm}(z,z^\prime) \epsilon_m(z^\prime) E_0(z^\prime)\,,
\label{secg}
\end{eqnarray}
where $$\tilde{D}_{nm}=D_{nm}(1-\delta_{n0})(1-\delta_{m0})$$ is  the differential operator restricted to the space of evanescent modes.

On the basis of Eqs.(\ref{extmod}) and (\ref{secg}) the formal correspondence with  quantum mechanical scattering processes in  many body systems is  established    explicitly by the following  substitutions (cf.\,\cite{FESH92})
\begin{eqnarray*}
\hspace{-0.0cm}D\to E-H, \;\; \tilde{D} \to Q( E-H)Q, \;\; D_{00}\to P( E-H)P,\,\;\;\epsilon_{n-0} \to -QVP,\quad \epsilon_{0-n} \to -PVQ\,.
\end{eqnarray*}
Here E denotes the total energy and  $H=T+V$  is the Hamilton operator of projectile plus target; $P$ and $Q$ are the projection operators on open ($E_0$) and closed channels ($\{E_n,\,n\neq 0\}$) respectively.

Relatives of the quantum mechanical bound states in the closed channels (eigenstates of $QHQ$) are the eigenfunctions of $\tilde{D}$  with vanishing eigenvalues.  They are the progenitors of the Feshbach resonances in photonic crystal slabs. Close to the frequency $\omega_0$ where one of the eigenvalues $\tilde{\eta}(\omega_0)=0$ vanishes,   $\tilde{D}^{-1}$ is dominated by the  corresponding (normalized)  eigenmode  $\tilde{\mathcal{E}}_{\tilde{\eta}}$  with eigenvalue $\tilde{\eta}(\omega)$  and, as in atomic or nuclear physics applications, can be approximated by
\begin{equation}
\tilde{D}^{-1}(z,z^\prime) \approx \frac{\tilde{\mathcal{E}}_{\tilde{\eta}}(z)\tilde{\mathcal{E}}_{\tilde{\eta}}(z^\prime)}{-\tilde{\eta}(\omega)}\,,\quad \int_{-\infty}^{\infty} dz\,\tilde{\mathcal{E}}^2_{\tilde{\eta}}(z)=1.
\label{redo1}
\end{equation}
We will refer to this  as resonance dominance approximation of an isolated resonance.   The coupling of  this eigenmode to  the extended mode is obtained by expressing the coupled evanescent modes $E_n$ in Eq.\,(\ref{extmod}) via Eq.(\ref{secg}) in terms of $E_0$ and $\tilde{\mathcal{E}}_{\tilde{\eta}}$. In this way the eigenmode acquires  a width and  a shift
 and  is turned into an isolated  ``Feshbach resonance''. This procedure is easily generalized to the case of two or more overlapping resonances where  two or more terms in the spectral representation of $\tilde{D}^{-1}$ have to be taken into account.

In scattering of light, by truncating the infinite dimensional  space of evanescent modes to one of a large but finite dimension,  practically  exact results can be obtained in numerical evaluations such as the ``RCWA method'' \cite{MOGA81} or ``transfer matrix'' techniques \cite{MASO08}.  With this in mind and in addition to resonance dominance,   we also   invoke truncation and compute the eigenfunctions $\tilde{\mathcal{E}}_{\tilde{\eta}}$ at the lowest non-trivial level,\,i.e.\,,
we replace  $\tilde{D}$ by its diagonal part
$$\tilde{D}_{nm} \approx D_{nn}\delta_{nm}(1-\delta_{n0})\,.$$ In this ``truncation approximation'' the exact eigenmodes  $\tilde{\mathcal{E}}_{\tilde{\eta}}$  with eigenvalues $\tilde{\eta}(\omega)$ are approximated   by  the   eigenmodes and eigenvalues $\mathcal{E}_\eta,\eta(\omega) $ of the diagonal elements $D_{nn}$  (cf.\,Eq.\,(\ref{syseq})).  Since for different $n$ these differ by constants  $-(k_x+n K_g)^2$ it is sufficient to consider the $n$-independent equation
\begin{equation}
 \Big(-\frac{d^2}{d z^2}
-\epsilon_0(z) \omega^2 \Big)\mathcal{E}_\eta(z)= \eta^E\, \mathcal{E}_\eta(z)\,.
\label{swvafc}
\end{equation}
The resonance condition is given by ${\eta}^E+(k_x+\nu K_g)^2=0$  with an appropriately chosen  integer $\nu$.
Within  resonance dominance and truncation approximation the coupled system of  extended  and single evanescent modes reads
\begin{eqnarray}
E_0(z)= E_0^{(+)}(z)- \omega^2\int dz^\prime g_0^E(z,z^\prime)\epsilon_{-\nu}(z^\prime) E_{\nu}(z^\prime)\,,\quad
E_{\nu}(z)=\frac{\omega^2 \mathcal{E}_{\eta}(z)}{\eta^E+(k_x+\nu K_g)^2}\int dz^\prime\mathcal{E}_{\eta}(z^\prime)  \epsilon_{\nu}(z^\prime) E_{0}(z^\prime).
\label{car2}
\end{eqnarray}
The 2nd equation has the form
\beq
E_\nu (z) = \sigma_\nu \,\mathcal{E}_{\eta}(z)\, ,
\label{propsdig}
\eeq
with  $\sigma_\nu$  measuring the degree of excitation of the guided mode.   Multiplying the first equation with $\mathcal{E}_{\eta}(z)  \epsilon_{\nu}(z)$ and  using the relation (\ref{propsdig}) yields after  integrating over $z$  a closed algebraic equation for $\sigma_\nu$.  Solving this  and using the  asymptotics of $E^{(+)}_0(z)$ (\ref{eq:boundcondTEE}) and the associated Green's function (\ref{grfcE})  one  obtains the expression for the total reflection amplitude. We record it for the most common case of piecewise constant index of refraction (Fig.\,\ref{dispn})   for which the components $\epsilon_{\nu}(z)$  are constants and  vanish outside the grating interval for $\nu\ne0$. We find
\begin{eqnarray}
r^E=r_0^E+\frac{i \sigma_\nu\epsilon_{-\nu}{\mathcal C}_+^E}{2 k_z}=r_0^E+\frac{i|\epsilon_\nu|^2\omega^4\mathcal{C}^{E\,2}_{+}/2k_z}{\eta^E+ (k_x+\nu K_g)^2+ |\epsilon_\nu|^2 \omega^4\Sigma^E}\,,
\label{tore}
\end{eqnarray}
with the background reflection amplitude $r_0^E$ (cf.\,Eq.\,(\ref{eq:boundcondTEE})), the coupling strengths $\mathcal{C}_{\pm}^E$ between extended and evanescent modes and the ``self - coupling'' $\Sigma^E$
\begin{eqnarray}
\mathcal{C}_{\pm}^E=\int_{I_g} dz \mathcal{E}_{\eta}(z) E_0^{(\pm)}(z)\,,
\quad
\Sigma^E= \int_{I_g} dz \int_{I_g} dz^\prime  \mathcal{E}_{\eta}(z)
g^E_0(z,z^\prime)   \mathcal{E}_{\eta}(z^\prime)\,.
\label{SigE}
\end{eqnarray}
$\Sigma^E$ is generated by transitions from the  guided to
the extended mode followed by  the propagation in the extended mode   and  the back transition to the evanescent mode.
The integrations are carried out over the grating interval ${I_g}$.
The real part of $\Sigma^E$  gives rise to a shift of the resonance position and its imaginary part to the width which accounts for the loss of intensity from the guided to the extended mode.  As for the exact solution also in resonance dominance this redistribution of intensity can be shown to preserve flux conservation leading to the standard relation  between reflection and transmission coefficients.

The formalism   developed so far is easily extended  to  TM polarized  light.  The starting point is  the wave equation for the y-component of the magnetic field
\begin{equation}
\label{TMW}
 -\Big(\partial_x\frac{1}{\epsilon(x,z)}\partial_x+\partial_z
\frac{1}{\epsilon(x,z)}\partial_z\Big) H(x,z)=\omega^2 H(x,z)\,.
\end{equation}
The Fourier transform with respect to the  x-coordinate converts  this wave equation into the coupled system   identical  to (\ref{syseq})  but for the Fourier components $H_m(z)$  of $H(x,z)$ and with the redefined  differential operators
\begin{equation}
D_{nm}=-\partial_z \gamma_{n-m}(z) \partial_z
 + (k_x+nK_g)(k_x+mK_g)\gamma_{n-m}(z)\,.
\label{DMnm}
\end{equation}
The coefficients   $\gamma_{n-m}(z)$  denote  either the   $(n-m)$-th Fourier components of $1/\epsilon(x,z)$ or the inverse of the Toeplitz matrix associated with $\epsilon(x,z)$ (cf.\,\cite{LLI96}). With this redefinition of $D_{nm}$, the formal development for TE and TM waves becomes identical.  We can proceed directly  as above applying resonance dominance and truncation approximations and obtaining  a coupled system  of  the extended and   resonating   evanescent modes (cf.\,(\ref{car2}))
\eqna
 H_0(z)=H_0^{(+)}(z)+\gamma_{-\nu}\int_{I_g} dz' g_0^H(z,z') \Theta_{0\nu}H_\nu(z')\,,  \quad
 H_\nu(z) = \frac{\gamma_\nu \mathcal{H}_{\nu}(z) }{\omega^2-\eta^{H}} \int_{I_g} dz^\prime  \mathcal{H}_{\nu}(z^\prime)\,
\Theta_{\nu0}H_0(z')\,.
\eqne
Here the differential operator
\beq
\Theta_{nm}=\overleftarrow{\partial_z}\;\overrightarrow{\partial_z}+(k_z+nK_g)(k_x+mK_g)\,,
\label{eq:defofThetanm}
\eeq
and the normalized eigenfunction, the evanescent mode,
\begin{equation}
 \Theta_{\nu\nu} \mathcal{H}_{\nu}(z) =  \eta^H\mathcal{H}_{\nu}(z)\,,\quad \int _{-\infty}^\infty dz\, \mathcal{H}^2_{\nu}(z) =1,
\label{eigfc}
\end{equation}
is chosen such that  its eigenvalue $\eta^H$ is close to $\omega^2$ -- the resonance condition in the TM case.
The magnetic fields $H_0^\pm(z)$ and the Green's function  $g_0^H(z,z')$ satisfy the homogeneous and inhomogeneous differential equation (\ref{EGR}) respectively with $D_{00}$ given by Eq.\,(\ref{DMnm}). The same boundary conditions for $H^\pm_0(z)$ are imposed as for $E^\pm_0(z)$ (Eq.\,(\ref{eq:boundcondTEE})).

The Green's function is given by
\begin{equation}
g_0^H(z,z^\prime) = \frac{ H_0^+(z_>)H_0^-(z_<)}{ -\gamma_0(z) w^H(z)}\,,
\label{grfcM}
\end{equation}
where $w^H(z)$ is the Wronskian given by the same formula as (\ref{grfcE}) with $E_0^\pm$ replaced by $H_0^\pm$. The factor $-\gamma_0(z)$  is due to the presence of the first order derivative in the differential operator $D_{00}$\,(\ref{DMnm}).  It makes the denominator z-independent as is most easily shown by applying Abel's formula (cf.\,\cite{KIBO03}).

The resulting reflection amplitude for the scattering reads
\begin{equation}
 r^H=r^H_{0}-\frac{i\sigma_{\nu}\gamma_{-\nu} \mathcal{C}_+^{H}}{2k_z} =r^H_{0}
-\frac{i|\gamma_\nu|^2 \mathcal{C}_+^{H\,2} /2k_z}{\omega^2-\eta^H-|\gamma_{\nu}|^2\Sigma^H}\,.
\label{TMr}
\end{equation}
The strength of the couplings $\mathcal{C}^{H}_{\pm}$  between guided and extended modes and the self coupling  $\Sigma^H$ are given by

\begin{eqnarray}
\label{sigh}
\mathcal{C}^{H}_\pm=\int_{I_g} dz \mathcal{H}_{\nu}(z)\Theta_{\nu0}H_0^{(\pm)}(z)\,,
\quad
\Sigma^H=\int_{I_g} dz \int_{I_g} dz' \mathcal{H}_{\nu}(z)\Theta_{\nu0}\,g_0^H(z,z')\,\Theta_{0\nu}^\prime\mathcal{H}_{\nu}(z')\,.
\end{eqnarray}
The  results (\ref{tore})   and (\ref{TMr}) determine  the TE and TM reflection amplitudes.  As was stressed in the Introduction   the quantities entering these expressions are not fitting parameters.  They are directly calculable
once  the properties of the GWS,\,i.e.,\,the dielectric constants  and the sizes of the various layers (cf.\,Fig.\ref{dispn}) as well as the wave vector of the incident light  are specified.
The  eigenvalues $\eta^{E}$ and $\eta^{H}$ and the corresponding eigenfunctions $\mathcal{E}_{\nu}(z)$ and $\mathcal{H}_{\nu}(z)$ (Eqs.\,(\ref{swvafc})\,and\,(\ref{eigfc})), the functions  $E_0^{(+)}(z)$ (\ref{EGR}) and correspondingly $H_0^{(+)}(z)$ together with  the corresponding Green's functions ((\ref{grfcE}),\,(\ref{grfcM}))  are solutions of simple one dimensional equations  with piecewise  constant coefficients. The background reflection coefficients $r_0^{E}$ and $r_0^{H}$  are straightforwardly extracted from  the asymptotics  $z\to -\infty$ of $E_0^{(+)}$  and $H_0^{(+)}$ respectively.
What   then remains to be determined are the  quantities  $\mathcal{C}_{+}^{E,H}$  and  $\Sigma^{E,H}$ which are obtained by  evaluation of   the integrals (\ref{SigE})  and (\ref{sigh}) respectively.
All  the  necessary calculations which we have outlined can be carried out analytically up to the determination of the eigenvalues $\eta^{E,H}$  -- the only quantities which  require numerical solution of  transcendental algebraic equations.

Let us  now ascertain the accuracy of our method.  We have chosen typical values for the parameters  of the GWS (cf.\,Fig.\,\ref{dispn}):   the thickness of grating and guided mode layers is 0.1 and 0.4 $\mu$m respectively,  the values of $\epsilon$ in   the superstrate,  the guided mode and  the substrate layer are 1, 4, and 2.25 respectively and  the  values of $\epsilon$ in the grating layer are 1 and 4, the grating period $\Lambda=0.87\,\mu$m and the duty cycle $d/\Lambda=0.5$.

\begin{figure}[ht]
\vskip -.01cm
\includegraphics[width=0.5\linewidth]{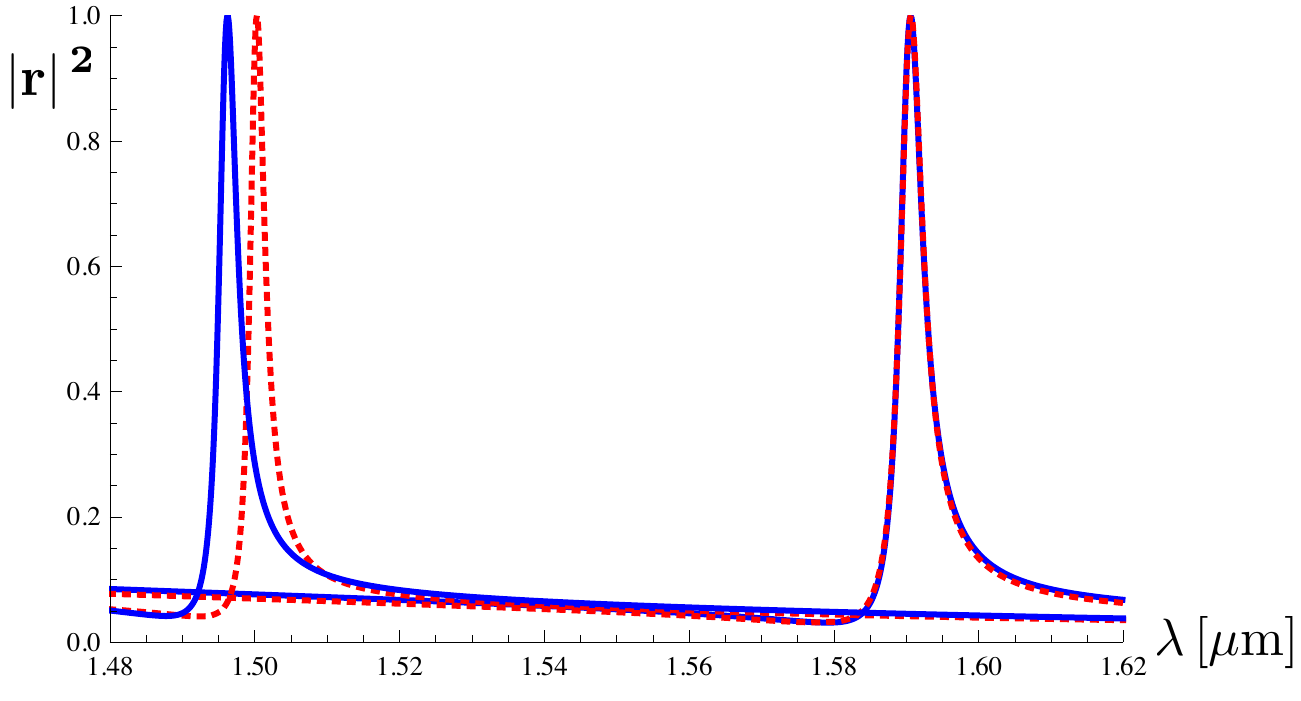}
\caption{Reflectivities as a function of the wavelength   for TM, Eq. (\ref{TMr}) (left peak) and TE, Eq. (\ref{tore}) polarized light incident at  $5^\circ$.   Solid line: exact results (obtained by applying transfer matrix techniques \cite{MASO08}),  dashed line: results of  resonance dominance and truncation approximations.}
\vskip -.06cm
\label{refc1}
\end{figure}

Fig.\,\ref{refc1}  displays the reflectivity of  the $\nu=-1$ TE and TM  resonances.
It is seen that already to lowest order in the truncation the  resonance dominance  yields a very good description of both the TE and TM resonances.  The  somewhat lower accuracy of the TM reflectivity reported    also in other approaches (cf.\,\cite{LLI96} and references therein) has its origin in  the derivative coupling (the second term in Eq.\,(\ref{TMW}))  between the modes. If applied to the  Green's function $g_0^H(z,z^\prime)$  the differential operator $\Theta_{\nu 0}$ (\ref{eq:defofThetanm}) generates a $\delta$ function which, in comparison with $\Sigma^E$(\ref{SigE}), makes    the self coupling $\Sigma^H$ (\ref{sigh})  more important and more sensitive to details of the guided modes.

\begin{figure}[ht] 
\includegraphics[width=0.5\linewidth]{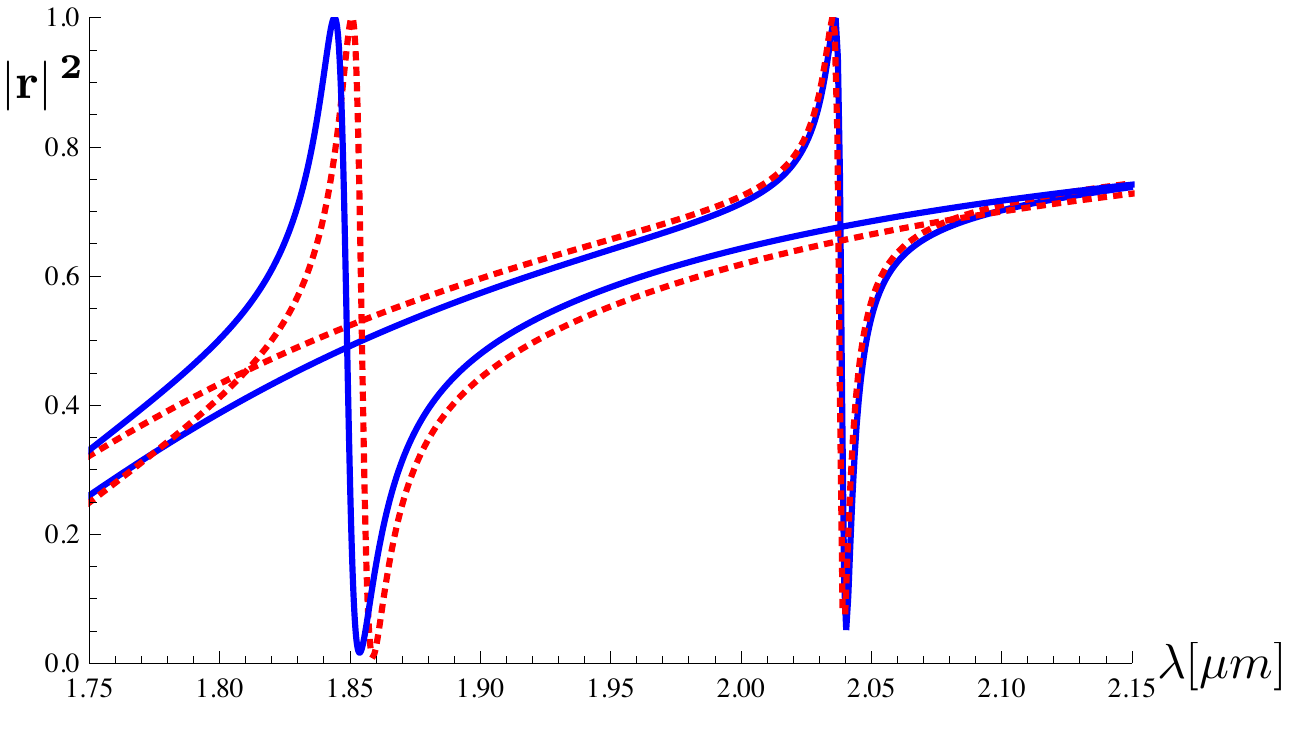}
\caption{ Same as in Fig.\,\ref{refc1} with period  $\Lambda=0.80\,\mu m$  for angles of the incident light of $85^\circ$ (TM) and $70^\circ$ (TE).}
\label{refc1M}
\end{figure}

A phenomenon that has received wide interest in various fields of physics is the  Fano interference \cite{FANO61},  also termed Fano or Fano-Feshbach resonances (cf.\,\cite{MFK10} and references therein). Its  distinct feature is the asymmetric shape of resonance curves.  In our context it finds its natural description in terms of interference of isolated Feshbach resonances and the background scattering ($r_0^{E,H}$) as indicated by the two terms in the reflection amplitudes in Eqs.\,(\ref{tore}) and (\ref{TMr}).
  The Feshbach resonance description of Fano interference  is closely related to the  ``temporal coupled-mode formalism'' of \cite{FSJ03} with  the background scattering corresponding to the ``direct pathway''.   We recall  that in our case the background scattering is generated for TE and TM polarizations by the effective dielectric medium defined by $\epsilon_0(z)$ and $\gamma_0(z)$ respectively.  It is negligible with the above choice of the  parameters  resulting in almost perfect Lorentzian shapes of the isolated resonances in Fig.\,\ref{refc1}. By changing the value of the period and of the incident angle we  can enforce  a strong background with a correspondingly strong  distortion of the resonance  shape as demonstrated in Fig.\,\ref{refc1M}. The level of agreement between exact and approximate calculations confirms  the validity of  the resonance dominance also in the presence of strong backgrounds.
\section{Overlapping  Feshbach resonances}  \label{ss2res}
The formulation of light scattering off GWS in terms of Feshbach resonances is readily generalized to the case of overlapping resonances. For  photonic crystal slabs of the structure shown in Fig.\,\ref{dispn}   overlapping resonances with both  TE and TM polarization   exist at illumination close to normal incidence independent of the detailed properties of the GWS.  For $k_x\to 0$    two guided modes are excited,\,i.e.,\,a resonance with $n=\nu$   is always  accompanied by the resonance with $n=-\nu$. We now present an analysis of this class of overlapping resonances.  Our treatment is easily generalized to the case of  two resonating
modes $n=\mu$ and $n=\nu$ with $\mu\ne -\nu$ (cf.\,\cite{EGLL11}).

 Accounting for the additional  mode in the resonance dominance approximation (\ref{redo1}) leads, after truncation, to the following system  of  coupled equations for TE waves
\begin{eqnarray}
\label{car21}
E_0(z) &=& E_0^{(+)}(z)  -\omega^2\int_{I_g} dz^\prime g^E_0(z,z^\prime)  \big(\epsilon_{-\nu}E_{\nu}(z^\prime)+\epsilon_{\nu}E_{-\nu}(z^\prime)\big)\,,\nonumber\\
E_{\pm \nu}(z) &=&  \frac{\omega^2}{\eta^E+(k_x\pm \nu K_g)^2}\,\mathcal{E}_{\eta}(z) \int_{I_g} dz^\prime\mathcal{E}_{\eta}(z^\prime) \big( \epsilon_{\pm \nu} E_{0}(z^\prime)+\epsilon_{\pm 2 \nu} E_{\pm\nu}(z^\prime)\big).
\label{car22}
\end{eqnarray}
In comparison with Eq.\,(\ref{car2}) this system  accounts for the coupling  of the extended to two guided modes as well as the interaction between the guided modes. In deriving it we have made use of the  property   that eigenvalue  $\eta^E$ and  eigenfunction  $\mathcal{E}_{\eta}(z)$ (Eq.\,(\ref{swvafc})) are independent of  $\nu$. Thus the resonating waves $E_{\pm \nu}(z)$  differ only in their (resonance enhanced)  strengths $\sigma_{\pm \nu}$ (cf.\,Eq.\,(\ref{propsdig})),\,i.\,e.,
\begin{equation}
 E_{\pm\nu}(z) = \sigma_{\pm\nu}\, \mathcal{E}_{\eta}(z)\,.
\label{sigma2}
\end{equation}
Multiplying the first equation of (\ref{car22}) with $\mathcal{E}_{\eta}(z)  \epsilon_{\pm\nu}(z)$ and  using  (\ref{sigma2}) yields after  integrating over $z$   two coupled  linear,\,algebraic  equations for  the  strengths parameters  $\sigma_{\pm \nu}$
\begin{equation}
W^E\left(
   \begin{array}{ccc}
   \sigma_{+\nu} \\
    \sigma_{-\nu} \\
   \end{array}\right)= \omega^2\, \mathcal{C}_{+}^E \left(
   \begin{array}{ccc}
   \epsilon_{\nu} \\
   \epsilon_{-\nu}  \\
   \end{array}\right)\,,
\label{sigsy}
\end{equation}
with the coupling matrix
\begin{eqnarray}
W^E=    \left(\begin{array}{ccc}
     \eta^E+(k_x+\nu K_g)^2+|\epsilon_{\nu}|^2 \omega^4\,\Sigma^E, & \hspace{-.5cm}\epsilon_{\nu}^{2}\omega^4\,\Sigma^E-\epsilon_{2\nu}\omega^2\,V^E \\
 \hspace{-1.4cm}\epsilon_{-\nu}^{2}\omega^4\,\Sigma^E -\epsilon_{-2\nu}\omega^2\,V^E\hspace{-.2cm}, & \hspace{-1.3cm}\eta^E+(k_x
-\nu K_g)^2+|\epsilon_{\nu}|^2\omega^4\,\Sigma^E
   \end{array}\right)\,.
\label{Msig}
\end{eqnarray}
The diagonal elements of $W^E$
contain the complex  valued self coupling  $\Sigma^E$ defined in Eq.\, (\ref{SigE}). In the  off-diagonal elements   direct couplings  $\epsilon_{\pm 2 \nu} V^E$  between the guided modes appear with
\begin{equation}
V^E=\int_{I_g} dz  \mathcal{E}^2_{\eta}(z)  \,,
\label{VE}
\end{equation}
 along with the indirect coupling terms $\epsilon_{\pm\nu}^{2}\Sigma^E$ via the extended mode.
In terms of the matrix elements  $W^E_{ij}$, the inverse of $W$ is given by
\begin{equation}
(W^{E})^{-1}=
    \frac{1}{W^E_+ W^E_-}\left(\begin{array}{ccc}
     W^E_{22} &- W^E_{12} \\
 -W^E_{21} & W^E_{11}\\
   \end{array}
\right)\,,
\end{equation}
with the eigenvalues of $W^E$
\begin{eqnarray}
&&\hspace{-.15cm}W^E_{\pm}= \eta^E +k_x^2+\nu^2K_g^2 +\big|\epsilon_{\nu}\big|^2\omega^4\, \Sigma^E
\pm
\Big[(2k_x\nu K_g)^2+ \omega^4\nonumber\\ &&\hspace{-.3cm}\,\cdot\Big(\big(|\epsilon_{\nu}|^2 \omega^2\,\Sigma^E \hspace{-.2cm}-|\epsilon_{2\nu}| \cos \varphi^E  V^E\big)^2\hspace{-.2cm} +|\epsilon_{2\nu}|^2 \sin^2 \varphi^E \, V^{E\,2}\Big)\Big]^{1/2}\hspace{-.2cm}.\nonumber\\
\label{solev}
\end{eqnarray}
We have introduced the relative phase between $\epsilon_{2\nu}$  and  $\epsilon_{\nu}^2$
\begin{equation}
e^{i\varphi^E}=\frac{\epsilon_{2\nu}\epsilon_\nu^{\star\,2}}{|\epsilon_{2\nu}\epsilon_\nu^2|}\,,
\end{equation}
which will be seen to distinguish
different types of interacting resonances.
The phase $\varphi^E$ appears as a result of the interference of the 2-step process via the extended mode and the one step process connecting directly the two guided modes.
The combination of the strengths $\sigma_{\pm \nu}$ which determines the reflection amplitude is easily calculated
\begin{eqnarray}
r^E&=&r^E_0+\frac{i}{2k_z} \big(\sigma_{\nu}\,\epsilon_{-\nu}+\sigma_{-\nu}\,\epsilon_{\nu}\big) \omega^2 \mathcal{C}^E_{+}
=r^E_0+\frac{i|\epsilon_\nu|^2\omega^4\mathcal{C}^{E\,2}_{+}W^E_0}{k_zW^E_+W^E_-}, \nonumber\\
W^E_0&=&
\eta^E+k_x^2+\nu^2K_g^2+  |\epsilon_{2\nu}|\cos  \varphi^E \,\omega^2 V^E \,.
\label{sieps}
\end{eqnarray}

The same procedure applies to the formulation of overlapping TM polarized Feshbach resonances. In order
to keep the presentation transparent we restrict it  to the limit of small $|k_x|$ where the interaction of the resonances is significant and where
the two guided modes
\begin{equation}
\mathcal{H}_{\eta}(z) = \mathcal{H}_{ \nu}(z)\approx \mathcal{H}_{- \nu}(z)\,,
\label{HH}
\end{equation}
can be identified. (The numerical results to be discussed later have been obtained without resorting to this approximation.) In this limit we obtain the following system of coupled equations for the two evanescent components $H_{\pm \nu}(z)$ and the extended mode $H_0(z)$
\begin{eqnarray}
H_0(z) &=&  H_0^{(+)}(z)+ \int_{I_g} dz' g^H(z,z') \left(\gamma_{-\nu} \Theta_{0,\nu} H_{\nu}(z') + \gamma_\nu\Theta_{0,-\nu} H_{-\nu}(z') \right),
\label{eq:TM3_H0}\nonumber \\
H_{\pm \nu}(z)&=& \frac{1}{\omega^2 - \eta^{H}} \mathcal{H}_{\eta}(z) \int_{I_g} dz' \mathcal{H}_{\eta}(z') \left(\gamma_{\pm\nu} \Theta_{\pm \nu,0} H_0(z') + \gamma_{\pm2\nu}\Theta_{\pm \nu,\mp \nu} H_{\mp\nu}(z') \right)\,,
\label{carr23}
\end{eqnarray}
which is of the same structure as the one for TE polarization and is solved in the same way.   We introduce the matrix ($W^H$) connecting the strength parameters and the Fourier coefficients of the (inverse) dielectric constant of the grating layer
\begin{equation}
W^H\left(
   \begin{array}{ccc}
   \sigma_{+\nu} \\
    \sigma_{-\nu} \\
   \end{array}\right)=  \mathcal{C}_{+}^H \left(
   \begin{array}{ccc}
   \gamma_{\nu} \\
   \gamma_{-\nu}  \\
   \end{array}\right)\,.
\label{sigsy2}
\end{equation}
Comparing  the systems  (\ref{car22}) and (\ref{carr23}) the matrix elements and the eigenvalues of $W^H$ can be read off from Eqs.\,(\ref{Msig}) and (\ref{solev})
\eqna
W^H_{\pm} = \omega^2 - \eta^H-|\gamma_{\nu}|^2\Sigma^H
  \mp \Big[(2\zeta k_x \nu K_g)^2+
\left(|\gamma_{\nu}|^2 \Sigma^H+|\gamma_{2\nu}|\cos{\varphi^H}V^H\right)^2 +|\gamma_{2\nu}|^2\sin^2{\varphi^H}V^{H\,2}\Big]^{1/2}\hspace{-.45cm}.\nonumber\\
\label{eq:rootsTM}
\eqne
For TM polarization,  the direct coupling via $\gamma_{\pm 2 \nu}$  between the two guided modes
is given by
\begin{equation}
V^H  =  \int_{I_g} dz\, \mathcal{H}_{\eta}(z) \Theta_{\nu,-\nu} \mathcal{H}_{\eta}(z)\,.
\label{VH}
\end{equation}
and the relative phase by
\begin{equation}
e^{i\varphi^H}=\frac{\gamma_{2\nu}\gamma_\nu^{\star\,2}}{|\gamma_{2\nu}\gamma_\nu^2|}\,.
\end{equation}
Furthermore, to lowest order,  we have taken into account the change of the eigenvalues $\eta^H$ for small $k_x$
\begin{equation*}
\Delta \eta^H_{\pm \nu}=\pm2\zeta \nu K_g k_x, \;\;\; \zeta=\int_{-\infty}^{\infty} dz \gamma_0(z)\mathcal{H}^2_{\eta}(z)\,.
\end{equation*}
The structure of the reflection  amplitude for TM polarization
\begin{eqnarray}
r^H= r_0^H - \frac{i|\gamma_{\nu}|^2 \mathcal{C}_+^{H\,2}W^H_0}{ k_z W^H_+W^H_-}\,,\quad
W^H_0=\omega^2 -  \eta^H+|\gamma_{2\nu}| \cos\varphi^H V^H\,,
\label{siepsH}
\end{eqnarray}
is the same as  that  for TE polarization (\ref{sieps}).

As for isolated resonances  our treatment of overlapping resonances has led us  to a well defined algorithm for the computation of the reflection coefficient  which does not contain any adjustable parameter.
The only new quantities appearing here are the direct interactions $V$ (Eqs.\,(\ref{VE}) and (\ref{VH})) which  again can  be evaluated  analytically in  closed form.
\section{Patterns of Interactions of Overlapping Resonances} \label{sec:interact}
Eqs.  (\ref{sieps}) and (\ref{siepsH}) provide a full solution for the overlapping resonance case. The analytic nature of these expressions allows  a detailed analysis of various patterns arising as the overlapping resonances affect each other. We start by observing that, as expected and independent of the   details,   the single resonance result (\ref{tore}) is reached in the limit where, for TE polarization,  the  square root in Eq.\,(\ref{solev}) is dominated by the term $\sim |k_x\nu K_g| $
\begin{eqnarray}
\frac{W^E_0}{W^E_+W^E_-}\hspace{.2cm} \longrightarrow_{_{\hspace{-.8cm}\large{|k_x|\to \infty}}}&&\sum_{\mu=\pm\nu}\frac{1}{2}\frac{1}{\eta^E+ (k_x+\mu K_g)^2+ |\epsilon_\mu|^2 \omega^4\Sigma^E}.\nonumber
\end{eqnarray}
This limit can also be shown to be satisfied  for  TM polarized resonances provided the approximation (\ref{HH}) is not invoked.

With decreasing $k_x$ the  coupling between the resonances sets in and modifies their properties. Independent of the strength of this coupling are the averages of  positions and  widths  of the  resonances
\begin{eqnarray}
\frac{W^E_++W^E_-}{2}= \eta^E +k_x^2+\nu^2K_g^2 +\big|\epsilon_{\nu}\big|^2\omega^4\, \Sigma^E\,,\quad
\frac{W^H_++W^H_-}{2} = \omega^2 - \eta^H-|\gamma_{\nu}|^2\Sigma^H \,. \nonumber
\end{eqnarray}
They  coincide with the   averages of  the  denominators of the reflection coefficients  (\ref{tore}) and (\ref{TMr}) for the isolated resonances with the indices $\pm \nu$.   Thus broadening of one of the resonances is accompanied by narrowing of the other.  This is  reminiscent  of motional narrowing \cite{BlPP48} or of  super- \cite{D54} and subradiance \cite{PCPCL85} for coupled emitters.

The coupling of the resonances increases or decreases their distance $\big|\mbox{Re}( W_+-W_-)\big|$
depending on the  structure of the GWS. As our results show, the change in the distance depends  on the interplay between strengths and  phase  of the self coupling $\Sigma$  and the direct coupling $V$.
  Obviously, for sufficiently large values of the direct coupling  terms  $|\epsilon_{2\nu}V^E|, |\gamma_{2\nu}V^H| $  one will find a repulsion of the resonances. We will discuss  an example of  such a case below. It is remarkable that the system of overlapping resonances can be tuned to exhibit either ``level'' repulsion or attraction by variation of an external parameter -  the wave vector  component $k_x$  of the incoming light.

A peculiarity of the interacting resonances is the appearance of a zero in the resonance amplitude,\,i.e.  when
$$
W_0(\omega,k_x)=0,
$$
in Eq. (\ref{sieps}).  The presence of this zero may distort significantly the shape of the reflectivity   a phenomenon reminiscent of Fano resonances. Indeed the expressions (\ref{sieps}) and (\ref{siepsH}) for  the reflection amplitudes have the structure of a product of a Lorentzian and of a Fano resonance.
The appearance of this zero can also be interpreted as a result of  a cancellation between the contributions from the  eigenstates of  $W^{E,H}$ (cf.\,Eqs.\,(\ref{sigsy}),\,(\ref{sigsy2})). This destructive interference causes a transparency window within the resonance and is therefore closely related to the ``EIT'' phenomenon (electromagnetically induced transparency), 
(cf.\,Refs. \cite{H97}\,,\cite{FLIM05}).
\begin{figure}[ht] \centering
\includegraphics[width=0.5\linewidth]{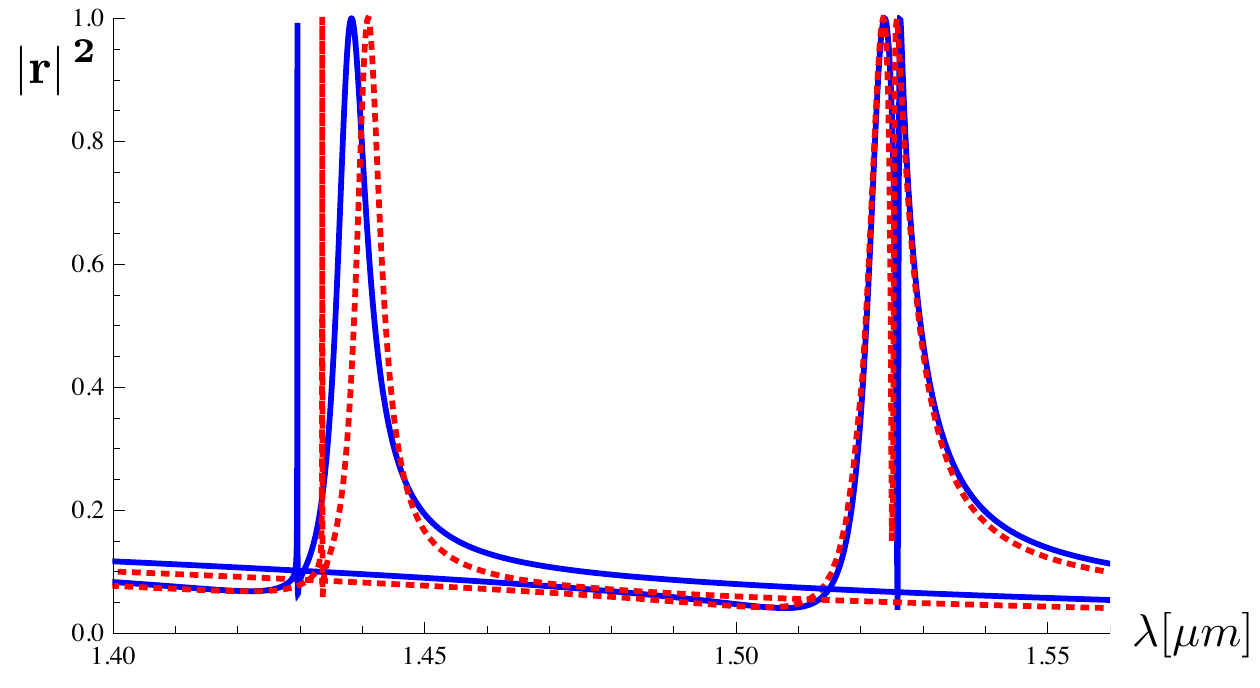}
\caption{Reflectivity for overlapping  resonances as a function of the wavelength for TM (left peaks,\,cf.\,Eq.\,(\ref{siepsH})) and  TE (Eq.\,(\ref{sieps})) polarized light incident at   $0.05^\circ$ with 50\% duty cycle.   Solid lines: exact results obtained by applying transfer matrix techniques, Ref. \cite{MASO08}.  Dashed lines: results of  resonance dominance and truncation approximations.}
\label{thresh2}
\vskip -.5cm
\end{figure}
The  limit of small $k_x$  offers further analytical insights into the relation between the dynamics and the structure of the interacting resonances.
We  consider first the case
\begin{equation}
\sin \varphi^{E,H} V^{E,H}=0.
\label{con1}
\end{equation}
As a function of the wavelength of the incident light the reflection coefficients for TE and TM polarizations exhibit   a single resonance for vanishing $k_x$ (cf.\,Eqs.\,(\ref{solev}),\,(\ref{eq:rootsTM})). Its width is twice as large as that of  an isolated  resonance (cf.\,Eqs.(\ref{tore}) and (\ref{TMr})).  In approaching the $k_x=0$ limit the widths of one of the  coupled resonances approaches zero. A singularity is prevented by  the presence of the  factor $W_0$ which, in this limit also approaches zero. In order to understand the transition from the single resonance  at $k_x=0$ to two interacting resonances  at finite but small $k_x$
we write  $W_{\pm}^E$ as
\begin{eqnarray}
&&\hspace{-0.25cm}W_+^E=  W_0^{E}(k_x) + 2(|\epsilon_\nu|^2 \omega^4\Sigma^E -|\epsilon_{2\nu}|\cos \varphi^E V^E)+\delta W^E\hspace{-.1cm},\nonumber \\
&&\hspace{-0.25cm}W_-^E =  W_0^{E}(k_x) -\delta W^E,\nonumber\\
&&\hspace{-0.25cm}\delta W^E = \frac{2(k_x\nu K_g)^2}{|\epsilon_\nu|^2 \omega^4\Sigma^E- |\epsilon_{2\nu}|\cos \varphi^E V^E}\,.
\label{kx0c}
\end{eqnarray}
For $k_x\neq 0$, the zero of $W_0$ in the numerator is no longer  canceled by the zero  of  $W_-$ which is shifted into the complex plane. At small but finite $k_x$ a narrow resonance is present together with  the zero of the numerator. The properties of the ``broad'' resonance ($W_+$) are only weekly affected. Whether or not  the  narrow resonance and the zero of $W_0$ overlap with the broad resonance, i.e.,\,the size of the bandgap, depends on the  shift
$
2(|\epsilon_\nu|^2 \omega^4\mbox{Re} \Sigma^E -|\epsilon_{2\nu}|\cos \varphi^E V^E)
$
which in turn is sensitive to the duty cycle.
In the limit considered the ratio $W_0^E/W_-^E$ generates a  Fano  resonance   and $1/W_+$ a Lorentzian.  Indeed expanding $W_-^E$ in the frequency around the zero of its real part yields the standard expression for Fano resonances \cite{MFK10}. The asymmetry (Fano-) parameter is given by $q=\cot \alpha$ with $\alpha$  denoting the phase
of  $|\epsilon_\nu|^2 \omega^4\Sigma^E- |\epsilon_{2\nu}|\cos \varphi^E V^E$.

The same analysis can be carried out for  overlapping resonances with TM polarization.
The shape of two TE and TM  overlapping resonances  calculated exactly and in  resonance dominance approximation are   shown in Fig.\,\ref{thresh2}. The results confirm  our analysis.  The coincidence of the narrow and wide resonances for TE polarization is due to the dominance  of the imaginary part of the self coupling $\Sigma^E$ (\ref{SigE}) in the expression for $\delta W^E$ (\ref{kx0c}) while their separation for TM polarization is a result of  the dominance of the real part of $\Sigma^H$  (\ref{sigh}).
\begin{figure}[ht] \centering
\vskip-.06cm
\includegraphics[width=.5\linewidth]{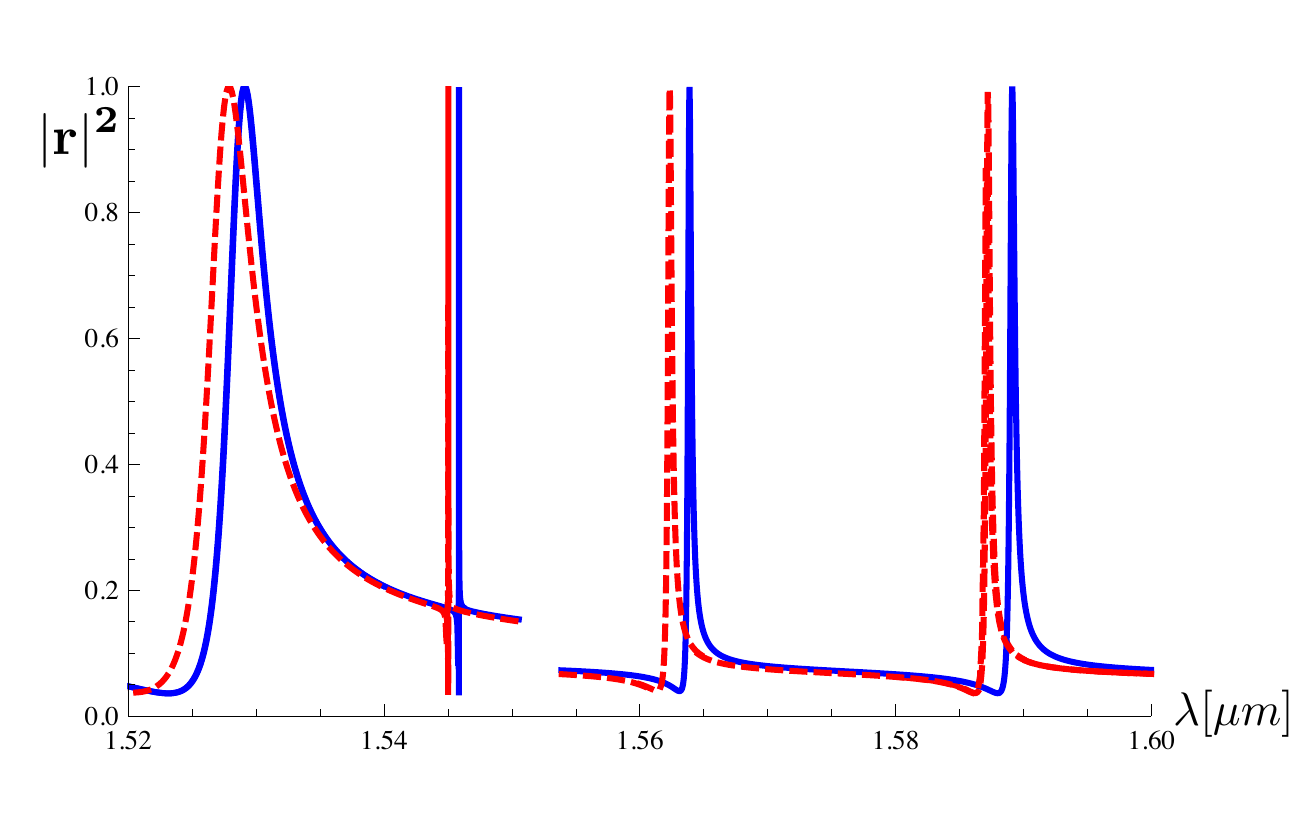}
\vskip-.02cm
\caption{Reflectivity for TE polarization as a function of the wavelength. Solid lines: exact values,  dashed lines: resonance dominance approximation. Left: The same structure and parameters as in Fig. \ref{thresh2} but with $75\%$ duty cycle. Right: $\theta=0.005^\circ$, with the grating structure (\ref{grser}) and the parameters (\ref{ergr}). The curves have been shifted to the right by 50 nm.}
\label{thresh3}
\end{figure}

Expression (\ref{kx0c}) for $\delta W^E$ indicates that the coincidence between the two resonances for TE polarization  does not occur  and the non vanishing gap between the resonances  is generated when  the direct coupling term $\epsilon_{2\nu}\,V^E$,  Eq. (\ref{VE}), is present. For symmetry reasons, the Fourier component $\epsilon_{2\nu}$ vanishes for 50\%  duty cycle. The left part  of Fig.\,\ref{thresh3} displays  the importance of the direct coupling term at 75\%  duty cycle. A band gap about 4 times as large as the width of the broad resonance is generated. Practically not  affected by this term is the ratio of about 500 between the broad and the narrow resonances.

A more balanced distribution of the widths can be achieved if
 the condition complementary to (\ref{con1})  is realized  for which the direct resonance  coupling  dominates the self coupling.
\begin{equation}
\varphi^{E,H}= \pm\frac{\pi}{2}\,,\quad \frac{|\epsilon_{\nu}^2| \omega^2|\Sigma^E|
}{|\epsilon_{2\nu}| V^E}\ll 1\,.
\end{equation}
In this limit we find (cf.\,Eq.\,(\ref{solev})) for sufficiently small  $k_x$
\begin{eqnarray}
W_{\pm}^E=
\eta^E +k_x^2+\nu^2K_g^2 +\big|\epsilon_{\nu}\big|^2\omega^4\, \Sigma^E \pm\Big[\omega^2 |\epsilon_{2\nu}|V^E+\frac{|\epsilon_{\nu}|^4\omega^8\Sigma^{E\,2}+4k_x^2\nu^2K_g^2}{2\omega^2|\epsilon_{2\nu}|  V^E}\Big]\,.\nonumber
\end{eqnarray}
The resulting  form  is quite different from what we have encountered so far.  The direct interaction  $V^E$ generates repulsion between the two resonances which guarantees that the  gap between the resonances does not vanish. The resonances  do not overlap and their widths are equal and coincide with   that of  an isolated resonance. Realization of this limit  can not be achieved  in  simple binary gratings such as the one  shown in Fig.\,\ref{dispn}. Let us instead consider the following more complicated structure
\beq
\epsilon(x)=\theta(\Lambda^2-4x^2)\Big[\epsilon_{\mbox{min}}+\delta\epsilon\, \big(\theta(-d-x)+\theta(x-d)\big)\Big],
\label{grser}
\eeq
with  the parameters
\beq
\Lambda= 0.87\,\mu m,\;\; d/\Lambda=0.2,\;\;\epsilon_{\mbox{min}}=1,\;\; \delta\epsilon =3\,.
\label{ergr}
\eeq
The corresponding results  are shown in the right part of  Fig.\,\ref{thresh3}  indicating that the desired limit is indeed realized in this structure  and confirming our analytical   insights.

Besides the reflection and transmission coefficients, our approach provides expressions for the  field strengths in particular in and close to the  waveguide layer (Fig.\,\ref{dispn}). For isolated resonances,  the strengths $\sigma_\nu$  of the evanescent modes  $E_\nu(z), H_\nu(z)$ are up to a non-resonating factor (cf.\,Eq.\,(\ref{tore})) given by the resonating part $r^{E,H}\hspace{-.1cm}-r_0^{E,H}$ of the corresponding reflection coefficient. For overlapping resonances a new resonating contribution to the field strengths arises. Using the equations (\ref {sigma2}-\ref{solev}) we evaluate  the evanescent electric field with the result
\begin{eqnarray*}
E_\nu(x,z) &=& \big(E_\nu(z)e^{i\nu K_g x}+E_{-\nu}(z)e^{-i\nu K_g x}\big)e^{ik_x x}\\
&=& \frac{2 \epsilon_\nu\omega^2 C_+^E e^{ik_x x}}{W^E_+ W^E_-}\Big\{W_0\cos \nu K_g x +\big(2 ik_x\nu K_g +|\epsilon_{2\nu}|\omega^2\sin \varphi^E V^E\big)\sin \nu K_g x\Big\}  \,\mathcal{E}_\eta(z)\,.
\label{fstr}
\end{eqnarray*}
As for isolated resonances, the first term is proportional to $r^E-r_0^E$, while the second term, due to the absence of $W_0$ in the numerator develops a singular point in the $\theta-\omega$ plane of the type
$$E_\nu(x,z)\sim \frac{\theta \tilde{\Sigma}}{\omega-\omega_0 +\theta^2\tilde{\Sigma}}\; ,$$
where $\tilde{\Sigma}$ is a complex constant.
In a lossless medium, at the frequency $\omega_0$  where $W_0^E$ vanishes, the evanescent contribution to the field diverges with vanishing angle of incidence $\theta$.
\section{Conclusion}
We have shown that resonances in light scattering off one dimensional photonic crystal slabs are, in a precise sense, Feshbach resonances.
This allowed us to developed a novel, accurate and essentially analytic  approximation scheme.   As demonstrated in the analysis of the intricate patterns of interacting resonances the advantage of our approach in comparison  with various types of parametrization  and exact numerical simulations resides in the explicit connection between  important quantities like the reflection coefficient and the structure of the GWS.
This  approach should  be useful   for   the  design of photonic crystal slabs with given requirements on reflection or transmission coefficient or on strength and phases of electric and magnetic fields. It   can be extended to the analysis of  several overlapping resonances or  to GWS built of layers of metamaterials,  cf.\,\cite{LZMH10}.  The extension  to two and three dimensional photonic systems appears to be feasible as well.
\vskip .3cm
F.L. is grateful for the support  and the hospitality at  the  Department of Condensed Matter, Weizmann Institute.  This work is supported in part by the Albert Einstein Minerva Center  for Theoretical Physics and by a grant from the Israeli Ministry of Science.


\begin{thebibliography}{0}
\bibitem{FESH58} H.\,Feshbach,\,{\em Ann.\,of\,Phys.}\,{\bf 5},\,(1958),\,357, {\em Ann. of  Phys.}\,{\bf 19},\,(1962)\,\,287.
\bibitem{FESH92} H.\,Feshbach, {\em Theoretical Nuclear Physics, Nuclear Reactions}, (John Wiley \& Sons,\,New York,\,1992)
\bibitem{TTHK99} E.\,Timmermans, P.\,Tommasini, M.\,Hussein, A.\,Kerman, {\em Phys.\,Rep.} {\bf 315},\,(1999),\,199.
\bibitem{CGJT09} C.\,Chin, R.\,Grimm, P.\,Julienne and E.\,Tiesinga,  {\em Rev. Mod. Phys.} {\bf 82},\,(2010),\,1225.
\bibitem{WARO907} S. S.\,Wang, R.\,Magnusson, J.S.\,Bagby, M.\,G.Moharam, {\em J. Opt. Soc. Am. A}\,{\bf 7},\,(1990),\,1470.
\bibitem{RSA97} D.\,Rosenblatt, A.\,Sharon, and A.~A.\,Friesem, {\em IEEE J. Quant. Electron.}\,{\bf 33},\,(1997)\,2038.
\bibitem{MP85}L.~Mashev and E.~Popov, {\em Optics Comm.} {\bf 55},\,(1985),\,377.
\bibitem{GSST85} G.~A.\,Golubenko, A.~S.\,Svakhin, V.~A.\,Sychugov, and A.~V.\,Tishchenko,
  {\em Sov. J. Quantum Electron.} {\bf 15},\,(1985)\,886.
\bibitem{KLGY10}
O.~Katz, J.~M.~Levitt, E.~Grinvald, and Y.~Silberberg, {\em Opt.\,Express} {\bf 18},\,(2010),\,22693.
\bibitem{WM93}
S.~S.~Wang and R.~Magnusson, {\em Appl.~Opt.} {\bf 32},\,(1993),\,2606.
\bibitem{MSJ10}
R.~Magnusson, M.~Shokooh-Saremi, and E.~G. Johnson, {\em Opt. Lett.} {\bf35},\,(2010)\,2472.
\bibitem{BFCBBDKTS10}
F.~{Br{\"u}ckner}, D.~{Friedrich}, T.~{Clausnitzer}, M.~{Britzger},
  O.~{Burmeister}, K.~{Danzmann}, E.~{Kley}, A.~{T{\"u}nnermann}, and
  R.~{Schnabel},  {\em Phys. Rev. Lett.} {\bf 104},\,(2010),\,163903.

\bibitem{FLPFB10}
D.~{Fattal}, J.~{Li}, Z.~{Peng}, M.~{Fiorentino}, and R.~G. {Beausoleil}, {\em Nature Photon.} {\bf 4},\,(2010),\,466.

\bibitem{MCWEC08}
M.~Lu, S.~S. Choi, C.~J. Wagner, J.~G. Eden, and B.~T. Cunningham, {\em Appl.\,Phys.\,Lett.} {\bf 92},\,(2008),\,261502.

\bibitem{DTZYJF00}  A.~{Donval}, J.~{Toussaere}, E.~{Zyss}, G.~{Levy-Yurista},E.~{Jonsson}, and A.A~{Friesem},  {\em Synthetic Metals} {\bf 124},\,(2001),\,19.



\bibitem{BLTFS09} O.~Boyko, F.~Lemarchand, A.~Talneau, A.-L.~Fehrembach, and A.~Sentenac,
  {\em J. Opt. Soc. Am. A} {\bf 26},\,(2009),\,676.

\bibitem{LTSYM98} Z.~S. {Liu}, S.~{Tibuleac}, D.~{Shin}, P.~P. {Young}, and R.~{Magnusson},
  {\em  Opt. Lett.} {\bf 23},\,(1998),\,1556.


\bibitem{KYFMHZ05} T.~{Katchalski}, G.~{Levy-Yurista}, A.~A. {Friesem}, G.~{Martin}, R.~{Hierle},
  and J.~{Zyss}, {\em Opt. Express} {\bf 13},\,(2005),\,4645.



\bibitem{FMS02} A.-L. {Fehrembach}, D.~{Maystre}, and A.~{Sentenac}, {\em J. Opt. Soc. Am. A} {\bf 19},\,(2002),\,1136.

\bibitem{FSJ03} S.~Fan, W.~Suh, and J.~D.~Joannopoulos, {\em J. Opt. Soc. Am. A},\,{\bf 20},\,(2003),\,569.

\bibitem{FAJO02} S.\,Fan, J.\,D.\,Joannopoulos, {\em Phys.\,Rev.B} {\bf 65},\,(2002),\,235112.

\bibitem{K03} K.~Koshino, {\em Phys.\,Rev.\,B} {\bf 67},\,(2003),\,165213.

\bibitem{MBPK95} M.\,G.\,Moharam,  D.\,A.\,Pommet, E.\,B.\,Grann, T.\,K.\,Gaylord, {\em J.\,Opt.\,Soc.\,Am.\,A} {\bf 12},\,(1995),\,1077.
\bibitem{SSS82} P.\,Sheng, R.\,S.\,Stepleman and P.\,N.\,Sanda, {\em Phys.\,Rev.\,B} {\bf 26},\,(1982),\,2907.
\bibitem{MFK10}
A.~E. Miroshnichenko, S.~Flach, and Y.~S. Kivshar, {\em Rev. Mod. Phys.} {\bf 82},\,(2010),\,2257.
\bibitem{M09} A.~E. {Miroshnichenko}, {\em Phys. Rev. E} {\bf 79},\,(2009),\,026611.


\bibitem{KCLG11}
Ming Kang, Hai-Xu~Cui, Yongnan~Li, Bing~Gu, Jing~Chen and Hui-Tian Wang
{\em J.\,Appl.\,Phys.} {\bf 109},\,(2011),\,014901.
\bibitem{ZHTH10} XU~DaZhi, LAN~Hou, SHI~Tao, DONG~Hui and SUN~ChangPu, {\em Science China} {\bf 53},\,(2010),\,1234.
\bibitem{EGLL11} I.\,Evenor, E.\,Grinvald, F.\,Lenz, S.\,Levit,  {\em arXiv:}1111.0208
\bibitem{JJWM08} J.\,D.\,Joannopoulos, S.\,G.\,Johnson, J.\,N.\,Winn and R.\,D.\,Meade, {\em Photonic Crystals Molding the Flow of Light}  (Princeton University Press,\,Princeton and Oxford,\,2008)
\bibitem{MOGA81} M.\,G.\,Moharam and T.\,K.\,Gaylord, {\em J.\,Opt.\,Soc.\,Am.} {\bf 71},\,(1981),\,811.
\bibitem{MASO08} P.\,Marko\v{s}, C.\,M.\,Soukoulis, {\em Wave Propagation},\,(Princeton University Press,\,Princeton and Oxford,\,2008)
\bibitem{MOFS53}P. M. Morse and H. Feshbach, {\em Methods of Theoretical Physics} Vol.\,1, (McGraw-Hill,\,New York,\,1953)
\bibitem{LLI96}  L.\,Li, {\em J. Opt. Soc. Am. A} {\bf 13},\,(1996),\,1870.
\bibitem{KIBO03} A.\,C.\,King, J.\,Billingham and S.\,R.\,Otto, {\em Differential Equations},\,(Cambridge University Press,\,Cambridge,\,2003)
\bibitem{FANO61} U. Fano, {\em Phys.\,Rev.} {\bf 124},\,(1961),\,1866.
\bibitem{BlPP48} N.\,Bloembergen, E.\,M.\,Purcell, and R.\,V.\,Pound, {\em Phys. Rev.} {\bf 73},\,(1948),\,679.

\bibitem{D54} R.~H.\,Dicke, {\em Phys. Rev.} {\bf 93},\,(1954),\,99.
\bibitem{PCPCL85} D.~Pavolini, A.~Crubellier, P.~Pillet, L.~Cabaret, and S.~Liberman, {\em Phys. Rev. Lett.} {\bf 54},\,(1985),\,1917.
\bibitem{H97} S.~E. Harris, {\em Physics Today} {\bf 50}, 36, (1997)
\bibitem{FLIM05} M.\,Fleischhauer, A.\,Imamoglu, J.\,P.\,Marangos, {\em Rev. Mod. Phys.} {\bf 77},\,(2005),\,633.
\bibitem{LZMH10} B.~Lukyanchuk, N.~I.~Zheludev, S.~A.~Maier, N.~J.~Halas, P.~Nordlander, H.~Giessen, and C.~T.~Chong, {\em Nat.\,Mater.} {\bf 9},\,(2010),\,707.
\end{thebibliography}
\end{document}